\documentclass[aps,pre,floatfix,nofootinbib,showpacs,twocolumn,groupedaddress]{revtex4}
\usepackage[sort&compress]{natbib}
\usepackage{graphicx}
\usepackage{epsfig}
\usepackage{color}

\begin{document}

\title{Non-equilibrium Lyapunov function and a fluctuation relation for stochastic systems:
Poisson representation approach}

\author{K.G. Petrosyan}

\author{Chin-Kun Hu}
\email{huck@phys.sinica.edu.tw}

\affiliation{Institute of Physics, Academia Sinica, Nankang, Taipei 11529, Taiwan}

\date{\today}

\begin{abstract}
We present a statistical physics framework for description of nonlinear non-equilibrium stochastic processes,
modeled via chemical master equation, in the weak-noise limit. Using the Poisson representation approach and
applying the large-deviation principle we first solve the master equation. Then we use the notion of the
non-equilibrium free energy to derive an integral fluctuation relation for nonlinear non-equilibrium systems
under feedback control. We point out that the free energy as well as some functionals can serve as non-equilibrium
Lyapunov function which has an important property to decay to its minimal value monotonously at all times.
The Poisson representation technique is illustrated via exact stochastic treatment of
biophysical processes, such as bacterial chemosensing and molecular evolution.
\end{abstract}

\maketitle

\section{Introduction}

Investigation of nonlinear non-equilibrium systems in science and technology has attracted
much attention in past decades \cite{prigogine, haken, stratonovich, keizer,10MaHu,huang,11HungHu}.
Among the challenges in this interdisciplinary field is to construct a unified theory that would be able to
describe various phenomena associated with these complex systems.

Many of the nonlinear non-equilibrium systems are modeled via a master equation \cite{85HuG,00QianH,02QianH,06QianH}, sometimes called the
chemical master equation \cite{vankampen,gillespie} although it can describe processes that span from (bio)chemical
reactions to ecological, epidemiological and evolutionary interactions of reagents involved. In this paper
we consider the master equation and apply the Poisson representation invented by Gardiner and Chaturvedi \cite{statphys,gardiner}
in an analogy with Glauber-Sudarshan P-representation technique for quantum master equations \cite{zoller}.
This is a powerful analytical technique that allows to derive exactly stochastic differential equations
in an auxiliary variable space and then obtain analytical results for the actual quantities via simple algebra.

Using the Poisson representation technique for the master equation we are able to construct a statistical
physics framework for nonlinear non-equilibrium systems via dealing directly with the equivalent set of
stochastic differential equations. In the weak noise limit it is possible to obtain probability distribution
functionals containing stochastic actions that describe time evolution of the underlying systems. Having
obtained these stochastic actions it then becomes customary to define a statistical physics or thermodynamics
description \cite{smith} of the non-equilibrium system under consideration. The fluctuation relations are
routinely derived for processes taking place with or without feedback control. We also show that for the
same case of weak noise the Poisson representation technique allows to obtain stochastic actions that possess
exact deterministic as well as multiplicative noise terms.

This statistical thermodynamic description opens up the possibility for obtaining some general principles
that may govern biophysical systems \cite{friston}. As a matter of fact, we reintroduce the notion of a free energy for
an arbitrary non-equilibrium state described by a probability distribution function, also called non-equilibrium
or information free energy \cite{gaveau,klimontovich,esposito,deffner}. A very important property of this quantity
is that the non-equilibrium free energy can serve as a Lyapunov function for both equilibrium and nonequilibrium states.

In order to illustrate the Poisson representation technique we consider biophysical processes,
governed by the master equation. First, we investigate the bacterial chemosensing described
by Monod-Wyman-Changeux allosteric model \cite{changeux} and show how the corresponding master equation
\cite{celani,tostevin} can be treated via the Poisson representation technique. We then point out
that the response and control experiments for biological sensory systems \cite{celani} can be treated
using the same approach. Second, we show that Crow-Kimura model \cite{crow} of neutral theory of
molecular evolution for finite populations can be studied with the use of the master equation introduced
in \cite{deem} (including horizontal gene transfer) and the Poissonian ansatz. Using the technique
it is straightforward to obtain dynamics equations for quasispecies populations. It is done via the
least action principle applied to the stochastic action potential derived via stochastic differential
equations that are equivalent to the master equation and have exact diffusion matrices for finite populations.

The paper is organized as follows. In Sec.II we review the Poisson representation technique that allows us
to convert the master equation to the equivalent set of stochastic differential equations. Sec.III presents
the procedure for obtaining the solution to the original master equation in the weak-noise limit. Lyapunov
functions for non-equilibrium states are introduced in Sec.IV where we also derive an integral fluctuation
relation that generally applies to stochastic processes under feedback control. Applications of the Poisson
representation technique are illustrated in Sec.V. Concluding Sec.VI sums up obtained results and outlines
possible directions of further research.

\section{The Poisson representation approach}

A variety of processes in science can be described via
the chemical master equation \cite{gillespie}. Systems with reactions among interacting
reagents are governed by the master equation for the probability distribution function of
the numbers of reagents involved. The Poisson representation technique for dealing with the
master equation has a few important advantages among other methods, e.g., the ability to work with
small number of interacting particles and time-dependent reactions rates \cite{statphys,gardiner}.
In this paper we will not consider these cases but will point out the advantages of this method in
proceeding with exact diffusion matrices via stochastic differential equations.

 In this section, we briefly  review the Poisson representation approach to the master
equation for the chemical reactions following \cite{statphys,gardiner,drummond}
and introduce relevant notations to be used in following sections.

Consider a set of $s$ (bio)chemical reactions with $n$ number of components
\begin{eqnarray}
\sum_i N^{j}_i X_i
\begin{array}{c}
k^+ _j\\
\rightleftharpoons\\
k^- _j
\end{array}
\sum_i M^{j}_i X_i
\label{set_react}
\end{eqnarray}
where $N^{j}_i$ and $M^{j}_i$ are the number of $X_i$ molecules in
the left- and righthand side of the reactions, $i=1,...,n$, $j=1,...,s$.
Consider that in the chemical reaction system, the total number of
 $X_i$ molecules is $x_i$. The master equation for the above set of equations is given by
\begin{eqnarray}
&&\partial_t P(\textbf{x},t) = \sum_j [t^- _j (\textbf{x} +
\textbf{r}^j) P(\textbf{x}+\textbf{r}^j,t) - t^+ _j (\textbf{x})
P(\textbf{x},t)] \nonumber \\ &&+ [t^+ _j
(\textbf{x}-\textbf{r}^j) P(\textbf{x}-\textbf{r}^j,t) - t^- _j
(\textbf{x}) P(\textbf{x},t)] \label{mastereq}
\end{eqnarray}
where $\textbf{x} = (x_1,x_2,...,x_n)$, $r^j _i = M^j _i - N^j _i$,
and reactions rates $t^\pm _j (\textbf{x})$ are
$t^+ _j (\textbf{x}) = k^+ _j \prod_i \frac{x_i !}{(x_i - N^j _i)!}$
and $t^- _j (\textbf{x}) = k^- _j \prod_i \frac{x_i !}{(x_i - M^j _i)!}$.

The Poisson representation for the solution $P(\textbf{x},t)$ of the master
equation can be expressed as
\begin{eqnarray}
P(\textbf{x},t) = \int d\alpha \prod_i \frac{e^{-\alpha_i}\alpha_i^{x_i}}{x_i !} f(\alpha,t)
\label{poisson}
\end{eqnarray}
where $\alpha=(\alpha_1,\alpha_2,...,\alpha_n)$
are the auxiliary variables that can take both real as well as complex values \cite{statphys,gardiner}.
Substituting this into the master equation (\ref{mastereq}) one obtains
\begin{eqnarray}
\frac{\partial}{\partial t} f(\alpha,t) &=& \sum_j \left[\prod_i
\left(1-\frac{\partial}{\partial \alpha_i}\right)^{M^j _i}
- \prod_i \left(1-\frac{\partial}{\partial \alpha_i}\right)^{N^j _i}\right] \nonumber \\
&\times& \left[k^{+} _j \prod_i \alpha^{N^j _i} _i - k^{-} _j
\prod_i \alpha^{M^j _i} _i\right] f(\alpha,t). \label{kramersmoyal}
\end{eqnarray}
$f(\alpha,t)$ is not a probability distribution function \cite{gardiner}. The relationship
between momenta of particle number $x$
distribution and the momenta/correlators of $\alpha$ is the following
\begin{eqnarray}
\langle x(x-1)...(x-n+1) \rangle = \int d\alpha \alpha^n f(\alpha) \equiv \langle \alpha^n \rangle.
\end{eqnarray}
For instance, $\langle x \rangle = \langle \alpha \rangle$ and $\langle x(x-1) \rangle = \langle \alpha^2 \rangle$.
From the latter one obtains the variance of particle number to be
$\langle(x-\langle x \rangle)^2\rangle = \langle(\alpha-\langle \alpha \rangle)^2\rangle + \langle \alpha \rangle$ or,
equivalently, $\langle(x-\langle x \rangle)^2\rangle - \langle x \rangle = \langle(\alpha-\langle \alpha \rangle)^2\rangle$.

For monomolecular reactions $\sum_i M^j _i \leq 1$ and $\sum_i N^j _i \leq 1$,
and therefore the Eq.(\ref{kramersmoyal}) becomes a first-order
partial differential equation. The absence of the second- and higher-order derivatives
means that there is no noise term. Thus the distribution remains Poissonian. The equations
that describe such reactions are the following ordinary differential equations
\begin{eqnarray}
\dot \alpha = \sum_j r^j _i \left(k^+ _j \prod_i \alpha^{N^j _i} _i - k^- _j \prod_i \alpha^{M^j _i} _i\right).
\label{mono}
\end{eqnarray}
That is the above equations coincide with conventional deterministic rate equations which govern the dynamics of monomolecular reactions.

For the case of bimolecular reactions, meaning that $\sum_i M^j _i \leq 2$ and $\sum_i N^j _i \leq 2$, the Eq.(\ref{kramersmoyal})
becomes a Fokker-Planck equation (FPE)
\begin{eqnarray}
\frac{\partial}{\partial t} f(\alpha,t) = - \sum_i
\frac{\partial}{\partial \alpha_i}\left[A_i (J(\alpha))
f(\alpha,t)\right] \nonumber \\
+ \frac{1}{2}\sum_{ik}
\frac{\partial^2}{\partial \alpha_i \partial \alpha_k}
\left[B_{ik}(J(\alpha))f(\alpha,t)\right] \label{fpe}
\end{eqnarray}
with the drift matrix elements $A_i(J(\alpha)) = \sum_j r^j _i J_j (\alpha)$ and the diffusion matrix elements
$$B_{ik}(J(\alpha)) = \sum_j J_j (\alpha) (M^j _{i} M^j _k - N^j _{i} N^j _k - \delta_{ik} r^j _i),$$ where one has
$$J(\alpha) = k^+ _j \prod_i \alpha^{N^j _i} _i - k^- _j \prod_i \alpha^{M^j _i} _i.$$

If one makes transformation to the density-like variables $\rho_i
= \alpha_i / V$ then the FPE (\ref{fpe}) takes the following form
\begin{eqnarray}
\frac{\partial}{\partial t} f(\rho,t) = - \sum_i
\frac{\partial}{\partial \rho_i}\left[A_i (J(\rho))
f(\rho,t)\right] \nonumber \\
+ \frac{\epsilon}{2}\sum_{ik}
\frac{\partial^2}{\partial \rho_i \partial \rho_k}
\left[B_{ik}(J(\rho))f(\rho,t)\right] \label{fpe2}
\end{eqnarray}
where $\epsilon = V^{-1}$ and the drift and
diffusion matrix elements have the same form as above with
$\alpha$ replaced by $\rho$.

\section{Weak noise limit}

The stochastic differential equations for the multidimensional
system described by the FPE (\ref{fpe2})  take the form \cite{gardiner}
\begin{eqnarray}
\dot \rho^\nu = A^\nu (\rho) + g^\nu _i (\rho) \xi^i (t)
\end{eqnarray}
where $\rho=(\rho_1,\rho_2,...,\rho_N)$ is the state vector and $\delta$-correlated white noise $ \xi^i (t)$ is defined via
$\langle \xi^i (t) \rangle = 0$ and $\langle\xi^i (t) \xi^j (t')\rangle = \epsilon\delta^{ij} (t-t')$, $i,j=1,2,...,N$
and the multiplicative noise relates to the diffusion matrix as  \cite{gardiner}
$B^{\nu\mu}(\rho)=g^{\nu}_i(\rho) g^{\mu}_i(\rho)$.

In the weak noise limit (small $\epsilon$) \cite{graham} the large deviation principle \cite{smith,touchette} leads to
\begin{eqnarray}
f(\rho / \epsilon, t) \propto \exp\left[-\frac{S(\rho / \epsilon, t)}{\epsilon}\right]
\label{pdf0}
\end{eqnarray}
for the quasiprobability distribution function, where the effective action $S$ is governed by the Hamilton-Jacobi equation
\begin{eqnarray}
\frac{\partial S}{\partial t} + H(\rho,p) = 0
\end{eqnarray}
with the following Hamiltonian
\begin{eqnarray}
H(\rho,p)=\frac{1}{2}B^{\nu\mu}(\rho)p_{\mu}p_{\nu} + A^{\nu}(\rho)p_{\nu}.
\end{eqnarray}

The equations of motion for the dominant (most probable) trajectories $\rho(t)$ are
\begin{eqnarray}
\dot \rho^{\nu} = \frac{\partial H}{\partial p_{\nu}} = B^{\nu\mu}(\rho)p_{\mu} + A^{\nu}(\rho)
\end{eqnarray}
with the equations of motion for the auxiliary variables $\textbf{p}(t)$
\begin{eqnarray}
\dot p^{\nu} = -\frac{\partial H}{\partial \rho_{\nu}} = -\frac{1}{2}
\frac{\partial B^{\mu\lambda}(\rho)}{\partial \rho_{\nu}}p_{\mu}p_{\lambda}
+ \frac{\partial A^{\mu}(\rho)}{\partial \rho^{\nu}}p_{\mu}
\end{eqnarray}

Now after rewriting Eq.(\ref{poisson}) using the distribution function (\ref{pdf0}) as
\begin{eqnarray}
P(\textbf{x},t) = \int d\rho \prod_i \frac{e^{-\frac{\rho_i}{\epsilon}}
\left(\frac{\rho_i}{\epsilon}\right)^{x_i}}{x_i !} f(\rho / \epsilon, t)
\label{poisson2}
\end{eqnarray}
we can apply the following asymptotic formulae for Poisson distribution $P(\mu,n)
=\frac{e^{-\mu}\mu^n}{n!}$ with large mean number $\mu$ \cite{curtis}
\begin{eqnarray}
P(\mu,n)=\frac{1}{\sqrt{2\pi\mu}}e^{-\frac{(n-\mu)^2}{2\mu}}
\end{eqnarray}
As one can see this is a Gaussian distribution with the same values for the mean and variance.
Large mean value is the case in Eq.(\ref{poisson2}) since $\epsilon$ is small.
Thus we substitute the asymptotic expressions into that equation getting
$$P(\textbf{x},t) = \int d\rho \prod_i \frac{e^{-\frac{(x_i-\rho_i/\epsilon)^2}{2\rho_i/\epsilon}}}{\sqrt{2\pi\rho_i/\epsilon}}
f(\rho / \epsilon, t)$$
and then
$$P(\textbf{x},t) \propto \int d\rho \prod_i \frac{e^{-\frac{(x_i-\rho_i/\epsilon)^2}{2\rho_i / \epsilon}}}{\sqrt{2\pi\rho_i / \epsilon}}
\exp\left[-\frac{S(\rho / \epsilon, t)}{\epsilon}\right].$$
Having that rewritten as
$$P(\textbf{x},t) \propto \int d\rho \prod_i \frac{e^{-\frac{(\rho_i - \epsilon x_i)^2}{2\rho_i \epsilon}}}{\sqrt{2\pi\rho_i \epsilon}}
\exp\left[-\frac{S(\rho / \epsilon, t)}{\epsilon}\right]$$
we can now apply the asymptotic expression for the $\delta$-function in order to replace the Gaussians
$$P(\textbf{x},t)\propto\int d\rho \prod_i \delta(\rho_i - \epsilon x_i) \exp\left[-\frac{S(\rho / \epsilon, t)}{\epsilon}\right]$$
and then after integrating over the $\rho$ variables we eventually arrive at
\begin{eqnarray}
P(\textbf{x},t) \propto \exp\left[-\frac{S(\textbf{x}, t)}{\epsilon}\right].
\label{pdf}
\end{eqnarray}
This is a true probability function distribution that is the solution of the
chemical master equation (\ref{mastereq}), in the weak noise limit.  It contains
the stochastic action potential $S(\textbf{x},t)$. As was mentioned above,
the stochastic action satisfies the Hamilton-Jacobi equation
\begin{eqnarray}
\frac{\partial S(\textbf{x},t)}{\partial t} + H(\textbf{x},\textbf{p},\lambda_t) = 0
\label{hje}
\end{eqnarray}
with auxiliary canonical momenta $\textbf{p}=\partial S(\textbf{x},t) / \partial \textbf{x}$ defining the Hamiltonian
as a function of $\textbf{x}$, and $\textbf{p}$
\begin{eqnarray}
H(\textbf{x},\textbf{p},\lambda_t)=\frac{1}{2}B^{\nu\mu}(\textbf{x},\lambda_t)p_{\mu}p_{\nu} + A^{\nu}(\textbf{x},\lambda_t)p_{\nu}
\label{hamiltonian}
\end{eqnarray}
with the same drift $A(\textbf{x},\lambda_t)$ and diffusion $B(\textbf{x},\lambda_t)$ matrices as in the FPE (\ref{fpe}).
The existence of the stochastic effective action $S(\textbf{x},t)$ governed by the Hamilton-Jacobi equation implies the principle of least action,
the Maupertuis' principle, that states that the true path of a system is an extremum of the action functional \cite{goldstein}.
Here we introduced a time-varying external parameter $\lambda_t$ that would implement a feedback control. Let us remind that one of
the advantages of the Poisson representation method is that all the above relations hold for time-dependent rates $k^+ _j$ and $k^- _j$
for forward and backword reactions. We point out that this is a zero-order approximation in $\epsilon$.
Thus we neglect the higher order corrections due to weak noise.

\section{Non-equilibrium free energy as a Lyapunov function and an integral fluctuation relation}

The stochastic processes considered above are associated with an underlying
Hamiltonian dynamics (in the weak-noise limit). That allows to obtain integral fluctuation relations for these
nonlinear non-equilibrium systems. But before doing so let us first reintroduce
the {\it non-equilibrium free energy} $F$ defined as \cite{gaveau,klimontovich,esposito,deffner}
\begin{eqnarray}
F(\lambda)=F_0(\lambda)+\epsilon D(P(\textbf{x},\lambda)\|P_{st}(\textbf{x},\lambda))
\label{nefe}
\end{eqnarray}
where $F_0$ is the steady-state free energy and
\begin{eqnarray}
D(P(\textbf{x},\lambda)\|P_{st}(\textbf{x},\lambda))=\int d\textbf{x}
P(\textbf{x},\lambda)\ln\frac{P(\textbf{x},\lambda)}{P_{st}(\textbf{x},\lambda)}
\end{eqnarray}
is the relative (Kullback-Leibler) entropy \cite{kullback} between $P(\textbf{x},\lambda)$
and the steady-state distribution $P_{st}(\textbf{x},\lambda)$ corresponding
to the same parameter $\lambda$, and $\epsilon$ is the effective noise intensity.
As it was shown in \cite{deffner} (see also \cite{klimontovich})
the non-equilibrium free energy defined via (\ref{nefe})
can serve as a true Lyapunov function for non-equilibrium states  since it decays monotonously to its minimal value at all times
\begin{eqnarray}
\frac{d}{dt}F(\lambda) \leq 0.
\end{eqnarray}

Generalizing, we point out that any following functional
with a function $\Phi(z)$ having second derivative everywhere positive \cite{yaglom}
\begin{eqnarray}
L_\Phi [P(\textbf{x},t)] = \int dx P_{st} (\textbf{x},\lambda) \Phi \left(\frac{P(\textbf{x},\lambda)}{P_{st}(\textbf{x},\lambda)} \right)
\label{lyapunov}
\end{eqnarray}
would serve as a Lyapunov function for both equilibrium and non-equilibrium steady states with a stationary
probability distribution function $P_{st}(\textbf{x},\lambda)$. That is
\begin{eqnarray}
\frac{d}{dt}L_\Phi [P(\textbf{x},t)] \leq 0.
\end{eqnarray}
The case considered above corresponds to $\Phi(z)=z \ln z$.

Let us now consider a stochastic process governed via the chemical master equation (\ref{mastereq})
in the weak-noise limit (small $\epsilon$) under a feedback control being done via parameter(s) $\lambda_t$
that would be the time-dependent reactions rates.
Having defined the non-equilibrium free energy and its change in time $\Delta F = F(\lambda) - F_0(\lambda)$
we begin with the following expression
\begin{eqnarray}
\left\langle e^{-\frac{1}{\epsilon}(\Delta H - \Delta F)-I} \right\rangle
= \int dx_0 dy P(x_0) P(y|x) e^{-\frac{\Delta H - \Delta F}{\epsilon}-I} \nonumber
\end{eqnarray}
where we have defined $\Delta H = H(x_t,\lambda_t) - H(x_0,\lambda_0)$
and the mutual information $I(x,y) = \ln \frac{P(y|x)}{P(y)}$ \cite{cover_thomas};
$P(x_0)$ and $P(y|x)$ are the initial and conditional probability distribution functions.

Then in the way similar to \cite{takahiro} we obtain the following integral fluctuation relation (see Appendix A)
\begin{eqnarray}
\left\langle e^{-\frac{1}{\epsilon}(\Delta H - \Delta F)-I} \right\rangle = 1.
\label{ifr}
\end{eqnarray}
 Note that this integral fluctuation relation is different from the Jarzynski
equality \cite{jarzynski} which does not have a feedback control. Equation (\ref{ifr}) is
also different from the equation by Sagawa and Ueda \cite{sagawa} with the feedback control;
the later can be applied only to systems initially being at equilibrium. Equation (\ref{ifr})
applies to the processes governed via the master equation (\ref{mastereq}) in the weak
noise limit where we have a solution to that master equation, the probability distribution
function (\ref{pdf}), that contains a stochastic action which is a solution of the Hamiltonian-Jacobi
equation (\ref{hje}) with the Hamiltonian presented in Eq.(\ref{hamiltonian}).

\section{Application to molecular biophysical systems}

\subsection{Bacterial chemosensing}

In order to illustrate the Poisson representation technique let us consider the process
of chemosensing in bacteria that is described by Monod-Wyman-Changeux allosteric model
\cite{changeux} governed by the following master equation \cite{celani,tostevin}
\begin{eqnarray}
\dot P(n) = k_r [(1-a_{n-1})P(n-1)-(1-a_n)P(n)] \nonumber \\
+ k_b [a_{n+1}P(n+1)-a_n P(n)],
\label{allosteric}
\end{eqnarray}
where $n$ instead of $x$ is used to represent the total number of molecules (particles).

%% fig. 1
\begin{figure}
\includegraphics[width=1.05\linewidth,angle=0]{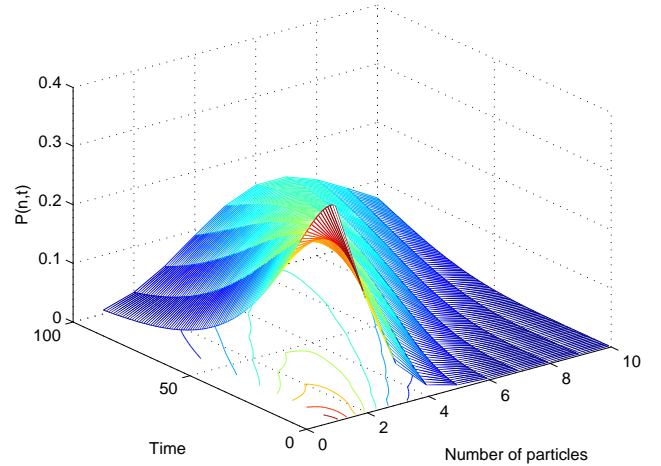}
\hfill
\caption{(Color online) Temporal evolution of the probability distribution
$P(n,t)$ as a function of the number of particles $n$ and time $t$ for
the process of bacterial chemosensing with the following set of parameters:
$N=6$, $n_0=1$, $\gamma=0.01$, $k_r=0.06$ $s^{-1}$, $k_b=0.12$ $s^{-1}$, $L=210 \mu M$,
$K_{on}=3000 \mu M$, $K_{off}=18.2 \mu M$.}
\label{fig1}
\end{figure}

%% fig. 2
\begin{figure}
\includegraphics[width=1.05\linewidth,angle=0]{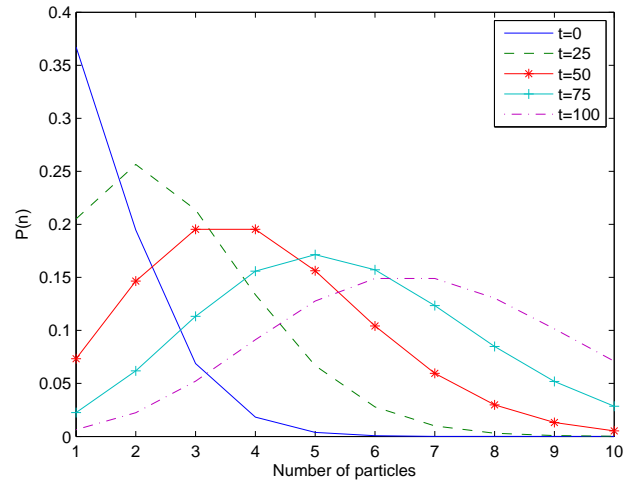}
\hfill
\caption{(Color online) Probability distribution function $P(n)\equiv P(n,t_i)$ at $t_i=0, 25, 50, 75$, and $100$
for the process of bacterial chemosensing with the same set of parameters as in Fig.(\ref{fig1}).}
\label{fig2}
\end{figure}

%% fig. 3
\begin{figure}
\includegraphics[width=1.05\linewidth,angle=0]{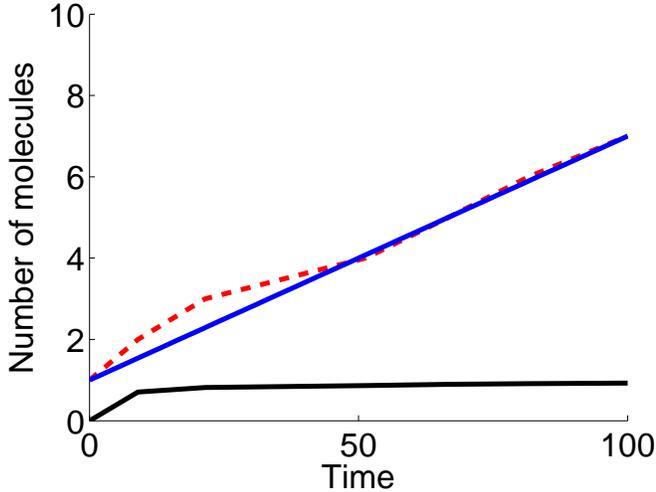}
\hfill
\caption{(Color online) Temporal evolution of the mean number of molecules $\bar n(t)$ (dashed (red) line)
and the Fano factor (the lower solid (dark) line), and the $\alpha$-variable (the solution of Eq.(\ref{ode}),
(the upper solid (blue) line) for the process of bacterial chemosensing with the same set of parameters
as in Fig.(\ref{fig1}).}
\label{fig3}
\end{figure}

The allosteric model describes a cluster of $N$ Tar receptor dimers with $M$ methylation sites.
Addition or removal of methyl groups take place at rates $k_r (1-a_n)$ or $k_b a_n$.
The cluster activity depends on the ligand concentration $L$ and the methylation state as
$a_n=\frac{1}{1+e^{f(n,L)}}$ with the free energy $f(n,L)=\gamma(n_0-n)+N\ln\frac{1+L/K_{off}}{1+L/K_{on}}$,
where $K_{on}$ and $K_{off}$ are the dissociation constants of ligand to the active and inactive receptors respectively.

Let us now apply the Poisson representation technique. The chemical master equation (\ref{allosteric})
describes a one-step reaction process and for slowly varying $a_n$ (provided $\gamma \ll 1$)
it is the case of monomolecular reactions (\ref{mono}).
The equation for the Poissonian $\alpha$ variable follows as
\begin{eqnarray}
\dot \alpha = k_r (1-a(\alpha)) + k_b a(\alpha).
\label{ode}
\end{eqnarray}
There is no noise term in this equation which means that the probability distribution function remains Poissonian
\begin{eqnarray}
P(n,t)=\frac{[\alpha(t)] ^n e^{-\alpha(t)}}{n!}
\label{me_solution}
\end{eqnarray}
with $\alpha(t)$ being the solution of the Eq.(\ref{ode}).  Figures (\ref{fig1}) and (\ref{fig2}) display the temporal evolution of the
probability distribution function.

In order to find out how well the Poisson distribution describes this stochastic process of bacterial chemosensing
we perform numerical simulations of the master equation (\ref{allosteric}) using the well-known Gillespie algorithm
\cite{gillespie_alg}.  Figure \ref{fig3} pictures temporal dynamics of the mean number of molecules $\bar n(t)$ and the
quantity $\frac{[n(t)-\bar n(t)]^2}{\bar n(t)}$ (the Fano factor) which should be equal to one for the Poisson distribution.
For comparison we also plot the time evolution of the solution of Eq.(\ref{ode}) for the $\alpha$-variable.
Taking into account that the number of particles is small (which means that the fluctuations could be large)
we conclude that, first, the statistics of the process rapidly approaches Poissonian and, second, the analytical
solution (\ref{me_solution}) to the master equation (\ref{allosteric}) describes the process quite well.

An open question though remains how to apply an integral fluctuation relation in order to investigate responses to
signals \cite{celani} as well as various cases of feedback control in the case of pure Poisson stochastic process.
We will leave that for further research stressing here that it was important to show that the model of bacterial
chemosensing could be efficiently studied using the Poisson representation technique.

\subsection{Fluctuations in molecular evolution}
Let us now consider an evolution of a finite population composed of $N$ binary purine/pyrimidine sequences, of length $L$.
We begin with the following master equation for the probability distribution $P({n_{\xi}},t)$ as a function of the set
of occupation numbers ${n_{\xi}}$ $(0\leq \xi \leq L)$ for the finite population Crow-Kimura model \cite{deem}
\begin{eqnarray}
\dot P(\{n_\xi\}) &=& \frac{1}{N} \sum_{\xi \neq \xi'} r(\xi)
[(n_\xi - 1)(n_{\xi'} + 1)P(n_\xi - 1, n_{\xi'} + 1) \nonumber \\
&-& n_\xi n_{\xi'} P(\{n_\xi\})] \nonumber \\
&+& \mu \sum^L _{\xi=0}
[(L-\xi)(n_{\xi}+1)P(n_{\xi}+1,n_{\xi+1}-1)
\nonumber \\&+& \xi(n_{\xi}+1)P(n_{\xi-1}-1,n_{\xi}+1) - Ln_{\xi}P(\{n_{\xi}\})] \nonumber \\
&+& \nu \sum^L _{\xi=0} [\rho_+
(L-\xi)(n_{\xi}+1)P(n_{\xi}+1,n_{\xi+1}-1) \nonumber \\ &+&
\xi\rho_- (n_{\xi}+1)P(n_{\xi-1}-1,n_{\xi}+1) \nonumber \\ &-&
n_{\xi}(\rho_+ (L-\xi) + \rho_- \xi)P(\{n_{\xi}\})],
\label{cme_evol}
\end{eqnarray}
where $r(\xi)=Lf(\xi)$ is the replication rate, $\mu$ is the rate of point mutation,
or of single base substitution, $\nu$ is the rate of horizontal gene transfer of single
letters between an individual sequence and the population. $\rho_{\pm} = (1 \pm u)/2$ is
the probability of inserting a wild-type or nonwild-type letter by horizontal gene transfer,
where $u = \frac{1}{N} \sum^L _{\xi = 0} (2\xi/L - 1)n_{\xi}$ is the "average base composition" \cite{deem}.

The master equation describes the following individual-based model of evolution
which includes the following replication, point mutation, and
horizontal gene transfer events
$\xi + \xi'
\begin{array}{c}
r(\xi)/N \\
\longrightarrow\\
\cdot
\end{array}
\xi + \xi$,
$\xi
\begin{array}{c}
\mu (L-\xi)\\
\longrightarrow\\
\mu \xi
\end{array}
\xi \pm 1$, and
$\xi
\begin{array}{c}
\nu \rho_+ (L-\xi)\\
\longrightarrow\\
\nu \rho_- \xi
\end{array}
\xi \pm 1$, respectively.

The FPE for the quasiprobability function $f(\{\alpha_\xi\},t)$ that is equivalent to the master equation (\ref{cme_evol})
in the Poisson representation reads as
\begin{eqnarray}
\dot f(\{\alpha_\xi\},t) = - \sum_\xi
\frac{\partial}{\partial \alpha_\xi}\left[A_\xi f(\{\alpha_\xi\},t)\right] \nonumber \\
+ \frac{1}{2}\sum_{\xi\xi'} \frac{\partial^2}{\partial \alpha_\xi \partial \alpha_\xi'}
\left[B_{\xi\xi'}f(\{\alpha_\xi\},t)\right]
\label{fpe_evol}
\end{eqnarray}
The deterministic term of the stochastic differential equations corresponding to the above FPE leads to
\begin{eqnarray}
\dot \alpha_\xi = A_\xi = \mu[(L-\xi+1)\alpha_{\xi-1}+(\xi+1)\alpha_{\xi+1}-L\alpha_\xi] \nonumber \\
+ \nu [\rho_+ (L-\xi+1)\alpha_{\xi-1} + \rho_-(\xi+1)\alpha_{\xi+1} \nonumber \\
-((L-\xi)\rho_+\xi\rho_-)\alpha_\xi] + [r(\xi)-\sum^L _{\xi'=0}r(\xi')\alpha_{\xi'}]\alpha_\xi \nonumber
\end{eqnarray}
that corresponds to the population level model for infinite population quasispecies theory \cite{deem}.
The fluctuation diffusion matrix can be obtained from the FPE (\ref{fpe_evol}) without the system size expansion
thus being exact for any finite $N$ (see Appendix B):
\begin{eqnarray}
B =
\left(
  \begin{array}{cc}
    \frac{2L}{N} f(\xi) \alpha_\xi \alpha_{\xi'} & -\frac{L}{N} [f(\xi)+f(\xi')] \alpha_\xi \alpha_{\xi'} \\
    -\frac{L}{N} [f(\xi)+f(\xi')] \alpha_\xi \alpha_{\xi'} & \frac{2L}{N} f(\xi') \alpha_\xi \alpha_{\xi'} \\
  \end{array}
\right) \nonumber
\end{eqnarray}

In the weak noise limit the above presented framework can be applied to the molecular evolution process
to obtain the probability distribution in the form of (\ref{pdf}) having the deterministic drift and the
fluctuation diffusion matrices variables substituted with species populations $n_\xi$. This, in turn, will
allow to get the integral fluctuation relation (\ref{ifr}) for the finite population Crow-Kimura model.
An integral fluctuation theorem was already applied to evolutionary processes \cite{lassig}. Our approach
makes it possible to study related processes for finite populations without system size expansion and in
time-varying environments or under some feedback control.

\section{Conclusion}

In summary, our paper contains following new results:

\begin{enumerate}

\item  It presents a novel framework to investigate various non-equilibrium
 nonlinear systems with a truly wide range of applications, including (bio)chemical
reactions, ecology, epidemiology, and molecular evolution.
 This approach allows to derive
 time-dependent probability distribution function for variables of finite open stochastic
 systems without a system-size approximation.

 \item It shows that a non-equilibrium
 free energy and some functionals can serve as Lyapunov functions,
 which has an important property to de-
cay to its minimal value monotonously at all times. The
next challenge might be to consider the non-equilibrium
Lyapunov function(s) to get some insight into various non-
linear stochastic (e.g., the evolutionary) processes.

 \item It contains a  novel, rather generalized, integral fluctuation relation (Eq.25)
 for nonlinear non-equilibrium systems including those under feedback control.

 Please note that our Eq. (25) is different from the equation considered by Jarzynski \cite{jarzynski}
 in which there is no feedback control. Our Eq. (25) is also different from the equation considered by
 Sagawa and Ueda \cite{sagawa}, which can be applied only to equilibrium systems.

 \end{enumerate}

 Our paper contain many new results and advance substantially the field of non-equilibrium statistical
 physics, especially small or finite open systems. The applications presented in the
 paper, the processes of bacterial chemosensing described by the Monod-Wyman-
Changeux allosteric model and the Crow-Kimura model of  molecular evolution,
 are in the focus of current research activities in leading world labs. The
 approach presented in the paper makes it much easier to deal with those problems.

We thank P. Grassberger for valuable discussions,  G. Hu, H. Orland and T. Sagawa for reading some parts
of the manuscript and useful comments, and M. Deem for useful conversations.
This work was supported by the National Science Council in Taiwan under Grant Nos. NSC 100-2112-M-001-003-MY2, NSC
101-2811-M-001-104, and NCTS in Taiwan.

\begin{appendix}
\section{Derivation of the integral fluctuation relation}
Let us derive the fluctuation relation (\ref{ifr}).
We begin with
\begin{eqnarray}
\left\langle e^{-\frac{1}{\epsilon}(\Delta H - \Delta F)-I} \right\rangle
= \int dx_0 dy P(x_0) P(y|x) e^{-\frac{\Delta H - \Delta F}{\epsilon}-I} \nonumber
\end{eqnarray}
where we have defined the change of the non-equilibrium free energy and of the Hamiltonian as $\Delta F = F(\lambda) - F_0(\lambda)$ and
$\Delta H = H(x_t,\lambda_t) - H(x_0,\lambda_0)$. The mutual information is $I(x,y) = \ln \frac{P(y|x)}{P(y)}$ \cite{cover_thomas};
$P(x_0)$ and $P(y|x)$ are the initial and conditional probability distribution functions.
Then we follow the lines presented in \cite{sagawa}
\begin{eqnarray}
\int dx_0 dy P(x_0) P(y|x) e^{-\frac{1}{\epsilon}(\Delta H - \Delta F)-I} \nonumber
\end{eqnarray}
\begin{eqnarray}
= \int dx_0 dy P(x_0) P(y|x) e^{-\frac{1}{\epsilon}(H(x_t,\lambda_t) - H(x_0,\lambda_0))}
e^{\frac{1}{\epsilon}\triangle F} \frac{P(y)}{P(y|x)} \nonumber
\end{eqnarray}
\begin{eqnarray}
= \int dx_0 dy P(x_0) P(y|x) e^{-\frac{1}{\epsilon}(H(x_t,\lambda_t) - H(x_0,\lambda_0))} \frac{Z_0}{Z_t} \frac{P(y)}{P(y|x)} \nonumber
\end{eqnarray}
%\begin{eqnarray}
%= \int dx_0 dy \frac{e^{-\frac{1}{\epsilon}H(x_0,\lambda_0)}}{Z_0} P(y|x)
%% e^{-\frac{1}{\epsilon}(H(x_t,\lambda_t) - H(x_0,\lambda_0))} %\frac{Z_0}{Z_t} \frac{P(y)}{P(y|x)} \nonumber
%\end{eqnarray}
\begin{eqnarray}
= \int dx_0 dy \frac{e^{-\frac{1}{\epsilon}H(x_t,\lambda_t)}}{Z_t} P(y) = \int dx_t \frac{e^{-\frac{1}{\epsilon} H(x_t,\lambda_t)}}{Z_t} = 1. \nonumber
\end{eqnarray}
The last calculation step used the Liouville's theorem $dx_0=dx_t$ \cite{goldstein}.
Thus we obtain the integral fluctuation relation
\begin{eqnarray}
\left\langle e^{-\frac{1}{\epsilon}(\Delta H - \Delta F)-I} \right\rangle = 1. \nonumber
\end{eqnarray}

\section{Fluctuation diffusion matrix for the finite population Crow-Kimura model}
The FPE (\ref{fpe_evol}) can be derived straightforwardly using the Eq. (\ref{kramersmoyal}) for the set of
"reactions", just as (\ref{set_react}), involving replication, point mutation and horizontal gene transfer
processes. It is easy to see that only the replication events $\xi + \xi'
\begin{array}{c}
r(\xi)/N \\
\longrightarrow\\
\cdot
\end{array}
\xi + \xi$
and
$\xi + \xi'
\begin{array}{c}
r(\xi')/N \\
\longrightarrow\\
\cdot
\end{array}
\xi' + \xi'$,
where $r(\xi)=Lf(\xi)$, contribute to the noise diffusion matrix elements $B_{\xi\xi'}$ as they involve
bi-reagent interactions. After some simple algebra without any approximation the $B$ matrix takes the
following exact form
\begin{eqnarray}
B =
\left(
  \begin{array}{cc}
    \frac{2L}{N} f(\xi) \alpha_\xi \alpha_{\xi'} & -\frac{L}{N} [f(\xi)+f(\xi')] \alpha_\xi \alpha_{\xi'} \\
    -\frac{L}{N} [f(\xi)+f(\xi')] \alpha_\xi \alpha_{\xi'} & \frac{2L}{N} f(\xi') \alpha_\xi \alpha_{\xi'} \\
  \end{array}
\right) \nonumber
\end{eqnarray}

\end{appendix}

\end{document}